\documentclass{article}

\usepackage{graphicx}

\begin{document}

\title{\textbf{Comments on an image encryption scheme based on a chaotic Tent map}}

\author{Alireza Jolfaei   \\             
 \small{Department of Computer and Information Sciences} \\
              \small{Temple University}\\
              \small{Philadelphia PA 19122}\\      
}



\date{}

\maketitle

\begin{abstract}
Recently an image encryption scheme based on a chaotic Tent map has been proposed by Li et al. \cite{Li:2016}. This comment shows that this scheme is broken and no secure application can be found for it.
\end{abstract}

\vspace{5mm}
The recent paper by Li et al. \cite{Li:2016} proposed a chaotic image encryption  based on a Tent map and evaluated its security through statistical analyses. It is true that when an image encryption scheme is designed, the security is generally evaluated through statistical analyses \cite{Jolfaei:2014,Jolfaei:2016}, and a cryptanalysis procedure is not commonly employed to assess image encryption schemes. However, before being applied to practical applications, where security is one of the important requirements, encryption schemes should be thoroughly analyzed using well-known cryptanalysis procedures. In this paper, it is shown that the image encryption scheme in \cite{Li:2016} is easily broken.

In a nutshell, as described in Section 3 of \cite{Li:2016}, the encryption algorithm is $c(i) = p(i) \oplus x(i)$, for $i = 1, 2, \dots, N$, where $c(i)$, $p(i)$, and $x(i)$ represent the $i$-th element of ciphertext, plaintext and expanded keystream, respectively; and $N$ is the number of plaintext elements. The expanded keystream is a sequence of pseudorandom numbers generated by a Tent map given by following equations:
\begin{equation}
x_i=\mu x_{i-1}, \ \mbox{if} \ \ x_{i-1} < 0.5,
\end{equation}
\begin{equation}
x_i=\mu (1 - x_{i-1}), \ \mbox{if} \ \ x_{i-1} \geq 0.5,
\end{equation}
where $\mu$ is a control parameter and $x_0$ is an initial seed. $(\mu, x_0)$ is considered as secret key. 

Firstly, given a plaintext with all-zero elements, the encryption algorithm discloses the the expanded key stream, that is, a sequence generated by a chaotic Tent map. As long as the key is not changed, the retrieved expanded keystream decrypts all ciphertexts correctly. 

Secondly, the confidentiality of the key relies on the security of a chaotic tent map which has already been analyzed by G. Alvarez et al. \cite{Alvarez:Cryptanalysis:2000}. In Sections 2 and 3 of \cite{Alvarez:Cryptanalysis:2000}, a detailed procedure is given to retrieve the seed point and the control parameter using recursive operations.

More importantly, the overall structure of Li et al.'s encryption algorithm is exactly similar to E. Alvarez et al.'s algorithm \cite{Alvarez:1999}, which has been cryptanalysed by G. Alvarez et al. in \cite{Alvarez:Cryptanalysis:2000}. It has been shown that such a cipher is broken with respect to ciphertext-only attacks, known/chosen plaintext attacks, and chosen ciphertext attacks.  

Following the above arguements, it is concluded that the encryption method of \cite{Li:2016} is easily broken and therefore cannot be considered at all as a serious encryption scheme.

\end{document}